\documentclass[prd,twocolumn,secnumarabic]{revtex4}
\usepackage{amssymb, amsmath}
\usepackage{graphicx}
\begin{document}

\newcommand{\lsim}{\,\raise 0.4ex\hbox{$<$}}
\newcommand{\weff}{w_\text{eff}}

\title{
Surveying the dark side}
\author{Roberto Trotta}
\email{rxt@astro.ox.ac.uk} \affiliation{Astrophysics Department,
University of Oxford, Denys Wilkinson Building, Keble Road, Oxford
OX1 3RH, UK}
\author{Richard Bower}
\email{r.g.bower@durham.ac.uk} \affiliation{Institute for
Computational Cosmology, Department of Physics, University of
Durham, South Road, Durham, DH1 3LE, UK}
\date{\today}

\begin{abstract}

We examine the prospects for the next generation of surveys aimed
at elucidating the nature of dark energy. We review the methods
that can be used to determine the redshift evolution of the dark
energy equation of state parameter $w$, highlighting their
respective strengths and potential weaknesses. All of the
attractive methods require surveys covering more than 5--10,000
sq.\ deg. of the sky. We examine the accuracy that each method is
likely to deliver within a decade, and discuss the difficulties
arising from systematic uncertainties associated with the
techniques.

We conclude that the proposed photometric and redshift surveys
have the potential of delivering measurements of $w$ with percent
accuracy at several redshifts out to $z \sim 3$. Of particular
interest will be the combination of weak lensing and baryonic
acoustic oscillations measurements. This exquisite precision is
likely to have a fundamental impact on our understanding of the
nature of dark energy, providing the necessary guidance for its
theoretical explanation.

\end{abstract}
\maketitle

\section{The Dark Sector of the Universe}

One of the most fundamental problems of contemporary physics is to
elucidate the nature of the ``Dark Sector'' of the Universe. A
wealth of cosmological observations seem presently to point to a
concordance cosmological model based on the Big Bang theory and an
homogeneous and isotropic Universe. Observations ranging from
measurements of temperature anisotropies in the cosmic microwave
background (CMB) to data on the abundance of hydrogen and other
light elements in the Universe indicate that  ``normal'' (i.e.\
baryonic) matter accounts for a mere 4\% of the matter--energy
contents of the cosmos. The remaining 96\% makes up the so--called
``Dark Sector'', with about 19\% of cold dark matter (CDM) and
77\% of ``dark energy''. The details of this cosmic budget vary
somewhat depending on the data sets used and the assumptions one
makes, but the errors on the different components are below 10\%
(see eg \cite{Spergel:2006hy,Sanchez} for details).

Direct evidence for the presence of non--baryonic, massive, cold
dark matter comes from observations of its gravitational effect on
eg the rotation curves of galaxies, or on the propagation of light
from distant sources (gravitational lensing). Further evidence
comes from the CMB, through which we now measure the epoch of
matter--radiation equality with better than 10\% accuracy. From
the theoretical point of view, there are many well--motivated
candidates for dark matter, for instance coming from
supersymmetric theories beyond the standard model of particle
physics. Direct and indirect detection experiments are now
starting to probe interesting regions of parameter space, and the
direct discovery of dark matter might well be within reach (see
\cite{deAustri:2006pe} for an example).

The situation is different as far as dark energy is concerned.
Observations of distant supernovae type Ia show that they apparent
luminosity is less than what one would expect in a
matter--dominated Universe
\cite{Tonry:2003zg,Perlmutter:1998np,Riess:1998cb}. This has been
interpreted as evidence for an accelerated expansion, that could
be caused by a new form of energy with negative pressure, dubbed
``dark energy'', whose energy density is about three times that of
matter at the present cosmic time \cite{Wang:2004py,Knop:2003iy}.
Similar conclusions can be reached by combining the CMB data with
observations of galaxy clustering (see eg \cite{Sanchez}). An
alternative explanation might lie in a change in our theory of
gravity, for instance by modifying General Relativity on
cosmological scales
\cite{Bekenstein:1984tv,Sanders:2002pf,Skordis:2005xk}.
Distinguishing between these two scenarios (ie, dark energy as a
modification to the equations of General Relativity or as the
manifestation of a new exotic source) is one of the main goals of
upcoming investigations.

\section{The nature of dark energy}

The difficulty comes from the fact that there is presently no
compelling theoretical explanation for the existence of dark
energy, nor for its present energy density. Guidance as to the
nature of dark energy has therefore to come from new observational
evidence. Only a better grasp on the properties of dark energy
will allow to solve outstanding questions such as the
``coincidence problem'', i.e.\ why is the proportion of dark
energy to dark matter of order 1 just now in cosmic history? Did
this happen simply by chance, or perhaps in virtue of some
anthropic principle, or because of some more fundamental, yet
unknown reason? There is also a  ``fine tuning'' problem, linked
with the small magnitude of the dark energy density, which lies at
least 60 orders of magnitude below theoretical expectations coming
from particle physics. From this point of view, the central
question is why is the energy scale of dark energy so small with
respect to other known scales of fundamental physics (the Planck
scale or the supersymmetry breaking scale)?

Advancements on all of those topics require stronger observational
proof of the properties of dark energy today and in the past. A
first important step is to discriminate between an evolving dark
energy (whose energy density changes with cosmic time) and a
cosmological constant of the form proposed by Einstein in the
1910s. A handle on this question is offered by the equation of
state parameter, $w$, that measures the ratio of pressure to
energy density of dark energy. If one could determine with high
accuracy that $w=-1$ and constant in time, this would strongly
support the case for a cosmological constant. This would imply
that dark energy is a manifestation of a new constant of Nature,
whose magnitude would however suffer from the above fine tuning
problem. Detecting an evolution with redshift of $w(z)$ would
support a dynamical form of dark energy, perhaps in the form of a
scalar field that could be linked to the inflationary phase of the
early Universe. Either one of these results is likely to have a
major impact on our knowledge of fundamental physics.

Current data are consistent with $w=-1$ out to a redshift of about
$1$. The uncertainty depends very much on the data sets used and
on the assumptions about the underlying model. If one models dark
energy in terms of $\weff$, an effective equation of state that is
constant but might be different from $-1$ (that can be understood
in terms of appropriate weighting functions depending on the
observable, see \cite{Simpson:2006bd}), then the uncertainty is of
order $5-10\%$, by using all of the available data sets (CMB, SNe,
Lyman-alpha absorption systems and large-scale structure)
\cite{Seljak:2004xh,Seljak:2006bg,Dick:2006ev}. However, one must
be very careful when assessing the combined constraining power of
different data sets whenever each one of them does not provide
strong constraints when taken alone. Combination of mutually
inconsistent data can potentially lead to unwarranted conclusions
on the dark energy parameters. This is why it is very important to
develop techniques that can determine $w$ accurately without the
need of combining many different measurements together.

A number of techniques can be used to investigate the nature of
dark energy, and they are reviewed in the next section.

\section{Probes of dark energy}

\label{sec:tech}

The observable impact of dark energy can be broadly divided in two
classes: modification of the redshift--distance relation and
effects on the growth of structures. Accordingly, we can divide
the different methods outlined below following the effect through
which they are mainly sensitive to dark energy: {\em probes of the
redshift--distance relation} (supernovae as standard candles,
acoustic oscillations as standard rulers) and {\em probes of the
growth of structures} (galaxy clustering, number counts, weak
lensing, Integrated Sachs--Wolfe effect).

We briefly describe each one in turn, highlighting their
respective advantages and weaknesses. Methods are listed in order
of approximately increasing number of assumptions about the
physical processes under observation, from the most ``clean''
(ISW, weak lensing and baryonic acoustic oscillations) to the ones
which require the most assumptions and are therefore less suited
to robustly probe dark energy.

{\bf The Integrated Sachs--Wolfe effect.} The transition from a
matter dominated to the dark energy dominated epoch induces a time
variation of the gravitational potentials, which in turn changes
the temperature of cosmic microwave background photons as they
traverse matter overdensities \cite{Sachs:1967er}. This is called
(late) integrated Sachs--Wolfe effect (ISW), and its measurement
requires the correlation of high--quality CMB maps with a large
survey of the galaxy distribution covering many thousands of
square degrees, under the assumption that light traces matter.

The ISW effect is a direct probe of the existence of dark energy,
but currently the significance of its detection is still low
\cite{Boughn:2004zm,Padmanabhan:2004fy}. The cross--correlation
measures the product of the growth function of density
perturbations and its time derivative averaged over a range of
redshifts, which is in principle a sensitive tracer of the dark
energy equation of state. However, the statistical power of the
method is limited by the fact that the signal is present only on
large scales and at low redshifts, where cosmic variance (the fact
that we have only one realization of the Universe to observe)
hinders the accuracy of the test
\cite{Pogosian:2004wa,Pogosian:2005ez}. The ISW effect is thus not
competitive with the statistical precision of the other methods to
determine a possible time--evolution of dark energy.

{\bf Weak gravitational lensing.} The distribution of matter in
the Universe causes gravitational bending of the light coming from
distant sources. The distortion pattern induced in the shape of
background galaxies and the change in their surface density
(magnification) by the intervening matter distribution can be
studied statistically. The measured signal is sensitive to a
combination of the geometry of the Universe, the growth and
distribution of structures, and the position of the sources (see
eg \cite{Schneider:2005ka} and references therein).

Information about the redshift distribution of the sources can be
exploited at different levels. Until now, the first generation
surveys have essentially measured 2--dimensional projections of
the lensing signal along the line--of--sight
\cite{Hoekstra:2005cs,Jarvis:2005ck}. The next generation surveys
will use photometric redshifts of the lensed galaxies to perform
an analysis in several redshift bins (``cosmic tomography''
\cite{Hu:2002rm}) to invert the lensing equations to deduce the
3--dimensional gravitational potential (``3D reconstruction''
\cite{Heavens:2003jx,Castro:2005bg}) or to carry out a geometric
test for dark energy \cite{Jain:2003tb} . Even a relatively coarse
division in 3 bins out to a redshift of $z=1$ can improve
constraints on the dark energy equation of state by a factor of 5
to 10 with respect to the 2D analysis.

The fundamental physics of weak lensing is well--known and the
method has the key advantage that it does not need to make any
assumption whatsoever about the mass--to--light relation. It is
therefore a very ``clean'' technique. This is potentially the
single most accurate method to study dark energy, but in order to
achieve the exquisite statistical accuracy it promises, a series
of systematic errors must be carefully controlled. Low seeing
(below 0.9 arcsec) is essential in order that accurate shapes can
be measured for sufficient numbers of distant background galaxies.
Since the weak lensing shear pattern presents distortions in the
shape of galaxies at the level of 1\% (0.1\%) on arcminute
(degree) scale, an accurate correction for instrument distortion
is required in order to extract the low amplitude cosmological
signal. The uncertainties in the photo--z calibration must be
controlled with exquisite accuracy (well below 1\% in random
errors and systematic bias), lest the tomographic advantage is
destroyed \cite{Ma:2005rc}. This is challenging but not unfeasible
and it requires large (of order of $10^4$) spectroscopic training
sets. An important contamination is intrinsic alignments between
galaxies, which can be mistaken for a cosmic shear signal.
Fortunately, photo--z information on the sources can be used to
down--weight pairs of galaxies which lie close in redshift,
thereby completely removing the intrinsic alignement of the
sources. Furthermore, the decomposition of the shear signal in
curl and gradient modes (called B and E modes, respectively)
allows for an important check of unsubtracted systematics, since
the weak lensing image distortions are curl--free to first order
\cite{Crittenden:2000au}. There is however another subtle effect
(gravitational--intrinsic correlations) which survives this
procedure, and which could potentially contaminate the signal at
the level of 10\%, without showing up in the B modes. Correlation
across different redshift bins could be used to decontaminate the
signal from this effect but this is a poorly understood systematic
effect and research is currently ongoing \cite{Hirata:2004gc}.
Mixing of the E and B modes due to lack of shear correlation
measurements is also a potential source of systematic error
\cite{Kilbinger:2006pr}.

{\bf Large--scale structures and acoustic oscillations.} From a
galaxy catalogue it is possible to extract the matter power
spectrum, given a model for the mass--to--light relation (bias).
The shape of the matter power spectrum probes mainly the matter to
radiation content and the spectral tilt of the primordial
fluctuations. When combined with CMB and/or SNe data, this yields
an indirect indication of the presence of dark energy, by
selecting a Universe with low mass density (see eg
\cite{Tegmark:2003uf}).

However, another approach based on a measure of standard rulers is
possible, since the tight coupling regime prior to recombination
imprints a characteristic oscillation scale in the power spectrum.
In the early Universe, gravitational compression of matter
overdensities is counterbalanced by radiation pressure in the
photon--baryon plasma. This results in acoustic waves that are
frozen ar recombination. Their characteristic scale, given by the
length sound could travel before recombination, has been imprinted
on the cosmic microwave background in the form of the oscillatory
peaks we now observe in the CMB power spectrum. This process also
leads to a preferred scale in the matter distribution,
corresponding to the acoustic horizon scale. ``Acoustic
oscillations'' or ``baryon wiggles'' appear in the correlation
function of galaxies as a bump that can be used as a standard
ruler for cosmological distance measurements
\cite{Eisenstein:1997ik,Blake:2003rh,Hu:2004kn}. The effect has
been recently detected in the correlation function of Luminous Red
Galaxies of the SDSS \cite{Eisenstein:2005su} as well as in the
2dF Galxy Redshift catalogue \cite{Cole:2005sx}.

At a given redshift, a measurement of the acoustic scale in the
direction transverse to the line of sight gives the angular
diameter distance, while the radial direction delivers a direct
measurement of the evolution of the Hubble parameter. Extracting
information on dark energy evolution requires two derivatives of
the former, but only one of the latter quantity. If the length
scale of the acoustic oscillations is known (eg, by calibrating it
against the observed peaks in the CMB angular power spectrum),
then the angular diameter distance and the Hubble rate are
measured absolutely by the above method. If such an absolute
calibration is not available, they are measured in units of the
present Hubble parameter, in which case the method needs to be
calibrated against a local measurement of the expansion rate. The
Alcock--Paczynski test is a weaker version of this procedure, in
that it does not determine the angular diameter distance and the
Hubble expansion separately, but only their product, by comparing
the extension of the standard ruler in the transverse and radial
directions at a given redshift.

To exploit fully the potential of acoustic oscillations,
spectroscopic redshifts are necessary to resolve the radial
direction, since photometric redshift information is not accurate
enough to measure the ruler in the redshift dimension. The washing
out of the radial direction means that a photometric survey must
cover about 20 times a larger area than a spectroscopic one to
achieve the same sensitivity to dark energy with the acoustic
oscillations technique \cite{Blake:2004tr}. This method requires
at the same time deep and wide surveys and is therefore very
expensive in terms of observing time. The advantage of the
technique is that the observation of the acoustic signature is
very robust to systematic errors, and non--linear effects, galaxy
biasing, redshift and lensing distortions can be corrected for
with good confidence \cite{Eisenstein:2004an,Eisenstein:2006nk},
even though further work is required to refine our understanding
of galaxy formation and evolution (\cite{White:2005tf}, but see
also \cite{Bower}). Furthermore, since the Hubble rate is the
derivative of the angular diameter distance, measuring both at the
same time provides an internal consistency cross--check. Another
key advantage is that measuring both quantities breaks a strong
degeneracy between the equation of state of dark energy and the
curvature of the Universe, allowing to distinguish between the
two.

{\bf Supernovae type Ia.} If the intrinsic luminosity of an object
and its redshift are known, measuring the flux at the observer's
position allows to reconstruct the luminosity distance to it. The
luminosity distance is an integral over redshift which depends on
the matter--energy content of the Universe. The presence of dark
energy in the form of a cosmological constant, for instance, makes
distant objects appear fainter by accelerating the expansion of
the Universe.

Light curves of stellar explosions known as supernovae type Ia
(SNe) can be empirically calibrated to give such a ``standard
candle'', and provided the first evidence of an accelerated
expansion and of the existence of dark energy in the late '90s
\cite{Perlmutter:1998np,Riess:1998cb}, that has been recently
confirmed by observations of supernovae at $z>1$
\cite{Riess:2004nr,Tonry:2003zg} . This method only measures ratio
of distances out to $z \lsim 1.5$ (see also \cite{Linder:2002bb}
for a discussion). It is therefore important that distant
supernovae can be empirically re--calibrated on local ones. The
danger with SNe methods is that the progenitors, and thus the
correction, may evolve systematically with redshift. This
potential systematic problem requires careful control through
spectroscopy. Future SN surveys will need to achieve better than
1\% control on SNe evolution. Large surveys detecting thousands of
SNe will allow a better control of this systematic by only
selecting similar galaxies at different redshifts. However, it
will be difficult to follow--up each SN spectroscopically, and
therefore it becomes important to be able to correctly select SNe
type Ia directly from the imaging survey, and to collect enough
information on their light curve to allow their recalibration. The
main systematics floor for this methods comes form our imperfect
knowledge of the details driving supernova Ia explosions.

{\bf Clusters number counts.} The value of the cosmological
parameters and dark energy properties can be constrained through
the cluster redshift distribution, which measures the number of
clusters per comoving volume per solid angle above a certain mass
threshold. The cluster mass can be estimated for instance through
the Sunyaev--Zel'dovich (SZ) effect, an increase in the
temperature of CMB photons produced by the rescattering of such
photons by the hot clusters gas they traverse. An alternative
method of selecting clusters would be using the galaxy colours,
e.g.\ the red sequence technique \cite{Galdders}. The approach of
combining this selection method with self--calibration may perform
as well as SZ--selection with the advantage that the clusters can
be selected directly from the imaging catalogue. The sensitivity
to dark energy comes in via the Hubble parameter, the angular
diameter distance and the mass selection function.

Unknown aspects of cluster physics might be an hindrance to
applications of this method for cosmology, and in particular for
dark energy studies. Of the methods proposed, this is arguably the
most susceptible to systematic errors induced by our imperfect
knowledge of cluster physics. The problem is that the abundance of
clusters is exponentially dependent on mass, and therefore a small
error in mass implies a large mistake in abundance. Thus, although
changes in the equation of state of dark energy result in large
variations at a given mass, the conversion between observable
quantities and cluster mass is highly uncertain (in both random
and systematic senses).

The success of the method will rely on ``self-calibration''
techniques that use the clustering of the clusters and the weak
lensing shear pattern around the cluster as additional constraints
\cite{Majumdar}. The ultimate precision that can be achieved with
these correction methods is not yet clear, and this is an area of
very active research.

\section{The next generation of surveys}

Ambitious observational campaigns targeting a combination of the
above techniques are being planned and will be carried out with a
new generation of instruments in the next decade. The field is
moving extremely fast, and funding agencies all over the world
(including PPARC, NASA, the US Department of Energy and ESO) have
given dark energy studies a considerable strategic priority. The
goals are on one hand to reduce the error on the dark energy
equation of state to the 1\% level, and on the other hand to
explore alternative explanations to the accelerated expansion.
Both of them will require a combination of the techniques outlined
above, but weak lensing and acoustic oscillations together appear
to be the most promising ones, both in terms of accuracy and for
maximising the discovery space.

\subsection{Imaging surveys}

Proposals for the next generation of imaging surveys driven by
dark energy science typically feature a survey area covering 5,000
to 10,000 square degrees, a large field of view (2 square degrees
or more) and four to five optical photometric bands. Those are the
basic specifications for both the {\em Dark Energy Survey} (DES,
see \cite{Abbott:2005bi}) and {\em darkCAM}, which would have
optical cameras mounted on 4m class telescopes. DES is a US--led
collaboration that will use a 520 megapixel CCD camera mounted on
the Blanco telescope to image 300 million galaxies at a median
redshift of $z \sim 0.7$ and to carry out weak lensing, baryonic
oscillations, cluster counts and SNe observations over 5 years,
starting in 2009. The UK involvement is led by UCL and has
recently been backed up by PPARC.  The European UK--led {\em
darkCAM} proposal to image some $10^9$ galaxies with weak lensing
image quality was originally envisaged to share time on ESO's
VISTA, but is now looking at a full--time site.

One of the most advanced projects is the {\em Pan-STARRS} survey
(Panoramic Survey Telescope and Rapid Response System,
\cite{panstarrs}), a US Air Force funded project in Hawaii,
primarily devoted to the identification of Earth-approaching
objects, but with 30\% of its time dedicated to supernovae, baryon
oscillations and weak lensing surveys. The first of the planned
four 1.8m telescopes is currently undergoing commissioning, and
the full system could be online by about 2009, representing a
major increase in power with respect to present--day surveys.

In purely statistical terms, the most precise constraints on the
dark energy equation of state are likely to come from weak
lensing. The details depend very much on which assumptions are
made about the cosmology and on which other data sets are
included. Another important factor is whether the analysis is
restricted to the safe linear regime or whether it includes
smaller scales modes, which give more aggressive constraints but
might suffer from less well controlled non--linear effects. On the
bright side, correlations among different observables can be
constructed from the survey, which will help improving both the
statistical accuracy by breaking parameter degeneracies and the
control on systematics: the correlation between foreground
galaxies number density (galaxy--galaxy correlation) and between
foreground galaxies and shear pattern (galaxy--shear
cross-correlation). Using all of this information, weak lensing
alone could achieve better than 5\% accuracy on the effective
equation of state, while in combination with CMB anisotropies
measurements of Planck quality (an ESA satellite missione due for
launch at the beginning of 2008 \cite{planck}) an accuracy of
1--2\% might be within reach.

This is of course only achievable if all of the systematic errors
will be kept closely under control. This means an exquisite image
quality, good seeing conditions (below 0.9 arcsec), excellent
photometric redshift reconstruction and control of intrinsic and
gravitational--intrinsic correlations. Arguably, the major
hindrance in pushing weak lensing constraints below the 5\% mark
will indeed come from systematic error control.

The clusters and SNe method will be considerably less stringent,
roughly a factor of 3 to 4 less precise than weak lensing, unless
combined with strong CMB priors (i.e., Planck data), in which they
case they will perform at about the 5\% level. The performance of
the cluster count technique relies however on self--calibration
using clustering and weak lensing data, a difficult procedure
compounded by the challenge of controlling systematic errors at
this level of precision. The possibility of SNe evolution and
missing pieces in our understanding of how a supernova explosion
comes about are also likely to be limiting factors when trying to
increase the accuracy on the equation of state below the 10--5\%
limit with this technique. Finally, measurements of acoustic
oscillations from imaging surveys are not competitive with the
other methods in terms of precision, reaching down to only about
20\% accuracy because of the lack of resolving power in the radial
direction (but see also \cite{Angulo}).

While none of the methods described above possess by itself all of
the {\em desiderata} that we would ideally want in trying to
constrain dark energy, combination of (at least) two techniques
offers many advantages. It allows for cross--calibration of
observables and facilites cross--checks of systematics, since the
physical underpinnings of each observable are different, and so is
the nature of the possible systematic errors. This last aspect is
very important in order to test the idea that dark energy is
indeed a new source in Einstein's equations (rather than e.g. the
manifestation of a different theory of gravity): by comparing
observables which are mainly sensitive to the growth of structures
with tests of the redshift--distance relation, we can look for
inconsistencies that cannot be explained by dark energy in the
form of a new fluid.

In view of this, a very promising combination is given by weak
lensing and baryonic acoustic oscillations, which together offer
the advantages of potentially high accuracy (weak lensing) and
robustness to systematics (acoustic oscillations). They
independently probe the growth of structures (lensing) and the
angular diameter distance relation (acoustic oscillations, once
calibrated against the high--redshift ruler given by the CMB). As
mentioned above, spectrographic redshift surveys encompassing
millions of galaxies will be needed to exploit fully the potential
of acoustic oscillations. We now turn to discuss the observational
perspectives in this field.

\subsection{Spectrographic surveys}

There are a number of redshift surveys at various stages of
planning, development or commissioning, that will have among their
main science drivers measurements of the acoustic ruler at
different redshifts.

Perhaps the most ambitious is the {\em Wide-Field Multi-Object
Spectrograph (WFMOS)} \cite{Bassett:2005kn}, a proposal for a 1.5
deg$^2$ multi-object spectrograph which will be able to observe
4,000 to 5,000 objects simultaneously. The instrument is to be
developed collaboratively by the Gemini and Subaru Observatories
and will be deployed on the 8m Subaru telescope on Mauna Kea,
Hawaii. Two baseline surveys are being proposed: a shallower and
wider one, covering 2,000 square degrees at $z~\sim 1$ which will
target emission line blue galaxies; and a deeper one, over 300
square degrees at $z \sim 3$ targeting Lyman-Break Galaxies. The
timescale for the deployment is 2013, with results over the first
500 square degrees due by the end of 2013 and completion of the
survey anticipated for 2016. Conceptual design studies are
currently being carried out.

The two WFMOS proposed baseline surveys will determine the angular
diameter distance and the Hubble expansion rate at $z \sim 1$ and
$z \sim 3$ with 1--2\% accuracy. The corresponding constraints on
the dark energy equation of state rely on the calibration of the
acoustic scale to Planck--like CMB observations and on an
independent measure of either the Hubble parameter or the dark
matter density parameter (eg via the matter power spectrum).
Combination with Planck forecasts and SDSS data gives an accuracy
in the range of 5--10\% in the effective equation of state. If one
drops the assumption of flatness, a constant $\weff$ can be
constrained to 5\% precision by combining WFMOS acoustic
oscillations measurements with SNe data, thus breaking an
important degeneracy between dark energy and spatial curvature.

On a shorter timescale, there are proposals to use the {\em
AAOmega} wide--field spectrograph -- an upgrade to the 2dF
spectrograph for the Ango--Australian Telescope, which has now
been successfully commissioned -- to carry out large surveys
(between 500 and 1,000 deg$^2$) in the redshift range $0.3 < z <
1$ to achieve 2\% accuracy in the angular diameter distance and
the expansion rate.

A rather more revolutionary concept is being investigated for the
{\em VIRUS} spectrograph, a proposal for the 9m Hobby-Eberly
Telescope in Texas based on industrial replication of low--cost
components \cite{Kelz:2006rs} . The aim is to measure $5 \cdot
10^5$ galaxies in a deep ($1.8 < z < 3.7$) and narrow (200
deg$^2$) survey by 2010, aiming to an accuracy of 1--2\% in the
diameter distance and the expansion rate. VIRUS is based on a
panoramic integral field spectrograph operated blindly without
target pre--selection and the technical feasibility of the survey
is being tested with a pilot project encompassing 10\% of the
modules.

In summary, the statistical accuracy from acoustic oscillations
redshift surveys is less than what could be achieved with weak
lensing. However, the acoustic oscillation method seems to be much
more robust with respect to systematic errors, and it can probe a
deeper redshift range than any other method. While a single
measurement of the acoustic scale at intermediate redshifts ($z <
1$) compared with the scale set by the CMB ($z \sim 1100$) gives
the largest lever arm for constraining dark energy, it would be
very interesting to measure the acoustic scale also in the deep
region, $z \sim 3$. This would allow to isolate unexpected
physical phenomena in the as yet unexplored range $3 < z < 1100$
(e.g.\ extra relativistic degrees of freedom), to observe the
deceleration epoch when dark energy is supposedly sub--dominant
and therefore check for an exotic dynamical behavior at $z>1$. A
deep measurement would also help in constraining deviations from
Einstein gravity that would not appear in tests based on the
growth of structures (see \cite{Yamamoto:2006yv} for an
illustration).

The WFMOS survey will require pre--selection of targets in the
desired redshift range, which should not represent a major
obstacle in the shallow region ($z\sim 1$), given that many
spectroscopic surveys (CFHT, Pan--STARRS, DES, VST--KIDS) should
provide the necessary coverage by 2012. The selection at high
redshift ($z \sim 3$) will require further effort to complement
the existing imaging data. A possible strategy is represented by
Hyper-Suprime, a proposal for a wide--field, 3 deg$^2$ camera for
the Subaru Telescope. This instrument would be able to produce a
high--quality weak lensing survey in 4 or 5 colours over 2,000
deg$^2$, which would be especially useful in conjunction with
WFMOS, covering the same region of the sky. Together,
Hyper-Suprime and \mbox{WFMOS} could exploit the complementarity
between weak lensing and acoustic oscillations, and WFMOS
spectroscopy would provide the necessary training sets for the
weak lensing tomography. Construction of Hyper-Suprime could
optimistically begin in 2009--2010, but the project is currently
only in the preliminary design phase and the funding status is
uncertain.

\subsection{Synergies between surveys and methods}

Dark energy science is a field that can benefit enormously by the
convergence of different observations, for the reasons recalled
above. In particular, the synergy with the next generation of CMB
experiment will be very important, either because this offers a
high--redshift calibration standard for acoustic rulers, or
because CMB data help reducing or breaking degeneracies between
dark energy and cosmological parameters.

Conversely, measurements of primordial B--modes (induced by
gravitational waves) in the CMB can profit from weak lensing
observations, by using them to subtract the contaminant signal
coming from line--of--sight lensing. For weak lensing, the
availability of IR photometric bands (for instance from the VISTA
IR survey) to complement visible photometry would yield
considerable improvement in the redshift accuracy for sources
beyond $z\sim1$, thereby helping in performing weak lensing
tomography.

Regarding acoustic oscillations, the most important limitation to
the absolute calibration is the accuracy with which the total
matter density is known. The present-day accuracy of 10\% is
insufficient for WFMOS, but it is expected that Planck will
improve this by a factor of 10, which would be enough to make this
uncertainty subdominant. For evolving models of dark energy and/or
a non--flat Universe, a precise knowledge of local standards
(i.e.\ the present Hubble parameter or matter density parameter)
would help to considerably tighten constraints. This can be
achieved with a variety of methods (SNe observations, shape of the
matter power spectrum, weak lensing, distance ladder measurements
of the Hubble parameter). The most useful single measurement is
probably SNe, since the degeneracy lines are oriented differently
in parameter space, and this would assists in breaking a
degeneracy between dark energy and spatial curvature.

Cluster selection and mass determination will be pursued by DES by
matching the area covered by the South Pole Telescope, a CMB
mission that is expected to measure some 20,000 clusters via the
SZ effect by 2009 \cite{spt}, another interesting example of
combination of techniques.

\subsection{A look into the future}

There is a series of proposals for next--to--next generation of
instruments, which aim at taking dark energy investigations to an
even more ambitious level.

The {\em LSST} (Large Synoptic Survey Telescope) is a project for
a wide--field, 8.4m telescope and a 3 Gpixels camera
\cite{Tyson:2002nh,Tyson:2003kb} . The survey will cover the whole
of the Southern hemisphere (or 20,000 deg$^2$) multiple times per
month with 6 colours photometry. It will survey the largest volume
ever proved, extending in the range $0.5 < z < 3$, and it will use
a variety of techniques (weak lensing, acoustic oscillations,
cluster abundance and a staggering 250,000 SNe per year) to
constrain dark energy at the percent level. The enormous volume
means that LSST is a potential competitor to spectroscopic surveys
such as WFMOS, as well, since the volume of LSST compensates for
the fact that its photometric redshifts will not be able to
resolve the radial mode in the acoustic oscillations. However,
this assumes that it will be possible to obtain precise
high--redshift photometric information without the use of
spectroscopic training sets, the prospects of which are uncertain.
The project is moving quickly and a very recent development saw
the award of substantial funding by the National Science
Foundation to carry on with the design and development stage. The
current schedule expects construction to begin in 2009 and first
light in 2013. Science will start in 2014. The entreprise is
technologically very challenging and a great amount of research
and development is still required to meet the specifications for
the telescope, the camera and the data analysis pipeline. Once
online, the LSST will quickly overtake all other surveys thanks to
its vastly superior survey speed.

The second half of the next decade will also see a great leap
forward in radio astronomy, as the SKA (Square Kilometer Array)
begins operations, first as a pathfinder (around 2015) and then as
a full system with a total collecting area of a million square
meters (around 2020) \cite{ska}. Thanks to its huge field of view,
the SKA will be able to measure redshifts of a billion of galaxy
over half of the sky in only a few months of operations, by
detecting radio emissions from hydrogen gas (see eg
\cite{Blake:2004pb}). The project is now beginning the design
study phase, thanks to a recent funding decision by the European
partners, including PPARC.

The possibility of a dark energy space mission remains uncertain
at the moment, with NASA and the US Department of Energy
reconsidering their strategic priorities regarding the concept of
a Joint Dark Energy Mission (JDEM), which is however unlikely to
be realized before the second half of the next decade.

\section{Summary}

Among all of the techniques reviewed, the most promising for dark
energy are weak lensing and acoustic oscillations, because of
their statistical accuracy (weak lensing) and robustness to
systematic errors (acoustic oscillations).

The weak lensing technique assumes only General Relativity and the
cosmological principle and it is based on well--known physics
(gravitational bending of light). The lensing signal is a
combination of the growth of structures, the cosmology and the
distribution of sources. Photometric redshifts are used to
reconstruct the radial dependence of the signal. Weak lensing
measurements can constrain the equation of state of dark energy
out to a redshift $z\sim 1$ to 1--2\% accuracy given a 5,000 to
10,000 square degrees, high--quality imaging survey in four to
five colour bands, with mild assumptions about the cosmology (eg
priors from CMB experiments). This represents a 5 to 10--fold
improvement over present--day constraints in this redshift range.
Achieving this precision requires an exquisite control of various
systematic errors, the dominant ones being photo--z accuracy,
image quality and intrinsic and galaxy--intrinsic correlations.
This calls for a factor of 10 improvement over current techniques
-- a difficult but not unfeasible task. There are several
in--built techniques which can be used to cross--check for
internal consistency and systematics.

Observations of baryonic oscillations with a spectroscopic survey
have less statistical power than weak lensing (roughly a factor of
5), but are less prone to systematic errors due to the
characteristics of the acoustic signature. The phenomenon of
acoustic oscillations is based on linear physics (at least at high
enough redshifts) that is well known from the CMB. The
oscillations' scale is used as a standard ruler to determine the
cosmology and is largely independent of the growth of structures.
The ruler needs to be absolutely calibrated to high--redshifts
standards (the CMB) and to local standards (the Hubble parameter
or matter density parameter). The method gives a complementary and
independent measurement of dark energy at intermediate, $z \sim 1$
redshift. It is also the only technique which can probe the
high--redshift evolution at $z \sim 3$, and check for consistency
in the dark energy evolution as inferred from lower--redshifts
probes.

It is also important to keep in mind that upcoming imaging surveys
such as DES, darkCAM or Pan-STARRS will produce a vast amount of
results in other fields apart from dark energy science, eg galaxy
and galaxy clustering evolution, star formation studies,
high--redshift quasars detection and evolution, local galaxy
studies, strong lensing, microlensing, near--earth objects and
outer solar system investigations, radio AGN's.

The WFMOS instrument will use spectroscopic redshifts of millions
of galaxies to detect the acoustic signature in the intermediate
and deep redshift range. When properly calibrated, this will allow
to reconstruct the redshift--distance relation and the expansion
rate with 1--2\% accuracy out to recombination time. With early
results expected for 2013--2014, WFMOS has the potential of
pioneering the field of wide and deep spectroscopic reconstruction
of the acoustic oscillations. Intermediate steps towards this
goals might be taken by the AAOmega spectrograph on a shorter
timescale.

In conclusion, the observational study of dark energy is a crucial
area of cosmological research. This is one of the most challenging
problems in contemporary physics, the solution of which is likely
to spark a new understanding of fundamental physics. Thanks to a
host of ambitious proposals and a strong support by several
funding bodies, key advances are likely to be made within the next
decade.

\section*{Acknowledgements}

RT is supported by the Royal Astronomical Society through the Sir
Norman Lockyer Fellowship and thanks PPARC for support of this
work. RGB thanks PPARC for the support of a Senior Research
Fellowship. The authors would like to thank all of the team
members of the DES, darkCAM and WFMOS collaborations for
stimulating discussions, and in particular Josh Frieman, Alan
Heavens, Ofer Lahav, John Peacock and Andy Taylor. RT would like
to thank Roger Davies and Joe Silk for useful comments. This work
is based on the {\em Dark Energy Review} commissioned to the
authors by the Science Committee of PPARC.

\end{document}